\documentstyle[12pt]{article}

\begin{document}
{} \hfill                                 Report No. IFT/8/97
\vskip1in

\centerline{\Large {\bf Asymptotic Freedom and Bound States}}
\vskip.1in
\centerline{\Large {\bf in Hamiltonian Dynamics}}
\vskip .2in
\centerline{ June 1997}
\vskip .4in
\centerline{\bf Stanis{\l}aw D. G{\l}azek}
\vskip .1in
\centerline{Institute of Theoretical Physics, Warsaw University}
\centerline{ul. Ho{\.z}a 69, 00-681 Warsaw}
\vskip.2in
\centerline{\bf Kenneth G. Wilson}
\vskip .1in
\centerline{Department of Physics, The Ohio State University}
\centerline{Columbus, Ohio 43210-1106}
\vskip.3in
\centerline{\bf Abstract}
\vskip.1in

We study a model of asymptotically free theories with bound states
using the similarity renormalization group for hamiltonians. We find 
that the renormalized effective hamiltonians can be approximated 
in a large range of widths by introducing similarity factors and 
the running coupling constant. This approximation loses accuracy for 
the small widths on the order of the bound state energy and it is 
improved by using the expansion in powers of the running coupling 
constant. The coupling constant for small widths is order 1. 
The small width effective hamiltonian is projected on a small subset 
of the effective basis states. The resulting small matrix is 
diagonalized and the exact bound state energy is obtained with accuracy 
of the order of 10\% using the first three terms in the expansion. We 
briefly describe options for improving the accuracy.


\vskip1.5in

\newpage

{\bf 1. INTRODUCTION}
\vskip.1in

So far, we do not have a precise theoretical description of the bound states 
of quarks and gluons in QCD which could simultaneously explain the parton 
model and the constituent quark model of hadronic structure. In particular, 
QCD is asymptotically free and the perturbative running coupling grows at 
small momentum transfers beyond limits. This rise invalidates the usual 
perturbative expansion in the region of scales where the bound states 
are formed. 

Ref.\cite{W3} suggested a light-front hamiltonian approach to this 
problem which is based on the calculation of the effective hamiltonians 
using the similarity renormalization group. \cite{GW1} \cite{GW2} Besides 
the calculation of the hadronic spectrum, one of our purposes is to 
obtain the quark bound state wave functions that can be used in the parton 
models of large momentum transfer processes. An alternative approach 
is the lattice gauge theory which is making progress but does not 
easily yield such wave functions. \cite{L} Other recent research
in the renormalized light-front hamiltonian approach to QCD is 
described in Ref. \cite{Perry}. 

Wegner \cite{WEG} proposed a flow equation for hamiltonians in solid 
state physics which is of the same kind as in the similarity renormalization 
group. Wegner's equation is based on an explicit form for the generator 
of the similarity transformation and corresponds to the Gaussian similarity 
factor with a uniform width.

This paper describes a numerical study of the key elements
of the hamiltonian approach in a simple matrix model which is 
asymptotically free, contains a bound state and can be diagonalized
exactly. We check the accuracy of different approximations in perturbation 
theory by comparison with the exact solution. The exact solution for
the renormalization group flow of the effective hamiltonians in the 
matrix model is obtained numerically using Wegner's equation. 

We start from the general assumption that the asymptotically free
theories have many degrees of freedom which are characterized 
by different scales of energy as measured by certain $H_0$. 
Then, we define matrix elements of the initial interaction 
hamiltonian $H_I$ between eigenstates of $H_0$. 

The model we study can be alternatively derived by discretization
of the 2-dimensional Schr\"odinger equation with a potential of 
the form a coupling constant times a $\delta$-function.\cite{J}
The continuous version of the 2-dimensional model has 
been studied by many authors. The discretized version was analyzed in Ref.
\cite{WG} using the exact solution to Wegner's equation. It was shown there
that the Wegner equation has the renormalization group interpretation.

This paper is organized as follows. In Section 2 we describe the model.
The parameters are chosen in a way which will make it clear that the method
of solution we use may have a wide range of other applications. Numerical 
results for the effective hamiltonians are presented in Section 3. We discuss 
the approximation based on the similarity factors and the running coupling
constant and we describe results obtained in perturbation theory. Section 4 
concludes the paper with a discussion of some options for improvements in 
perturbative calculations of the effective hamiltonians in the range of 
widths near the bound state formation scale.
\newpage
{\bf 2. MODEL}
\vskip.1in

The hamiltonian $H_0$ is assumed to have a finite discrete set of eigenstates,

$$ H_0 |i> \quad = \quad E_i |i> \,\, . \eqno(2.1) $$

\noindent The eigenstates are orthogonal and normalized,

$$ <i|j> \quad = \quad \delta^{ij} \,\, . \eqno(2.2) $$

\noindent The dynamics of states in the space spanned by this set
is defined by the interaction hamiltonian, $H_I$, whose matrix elements 
are assumed to be

$$ <i| H_I |j> \quad = \quad - \,\, g \, \sqrt{ E_i E_j } \,\, , \eqno(2.3) $$

\noindent where $g$ is a dimensionless coupling constant. 
The whole hamiltonian is denoted by $H$, $H = H_0 + H_I$.

In the current study, we choose the eigenvalues of $H_0$ in the form

$$ E_i \quad = \quad b^{2i} \,\, . \eqno(2.4) $$

\noindent In the numerical calculations we use $b = \sqrt{2}$. 
The integer power $i$ ranges from $M$ to $N$. The integer 
$M$ is considered to be large and negative. The lower, infrared bound on 
the free energies is $2^{-M}$. The integer $N$ is considered to be 
large and positive. The upper, ultraviolet free energy bound is given
by $2^N$. In the numerical study, we use $M=-21$ and $N=16$.

For the purpose of analogy to QCD, we adopt the convention that the 
energy equal 1 corresponds to 1 GeV, although the units of energy are 
arbitrary. Thus, the ultraviolet cutoff corresponds to 65 TeV and the
infrared cutoff corresponds to 0.5 eV.

With the above choices, the hamiltonian $H$ is a $38 \times 38$ matrix.
For the coupling constant $g > 1/38$, the matrix has one negative eigenvalue 
and 37 positive eigenvalues. The coupling constant is adjusted to obtain 
the negative eigenvalue equal $-1.00000000$ or, in our convention, $-1$ GeV. 
Namely, $ g = 0.06060600631$. The many digits are given for readers interested
in the numbers.

The eigenstate with the negative eigenvalue corresponds to
the s-wave bound state in the continuum Schr\"odinger equation
with the $\delta$-potential. The positive eigenvalue eigenstates 
correspond to the s-wave scattering states. We refer the 
reader to the work of Jackiw \cite{J} for details. The only new steps
required are the replacement of the continuous energy scale 
in Ref. \cite{J} by the discrete one in Eq. (2.4) and the introduction
of the infrared and ultraviolet cutoffs. These steps are described 
in Ref. \cite{WG}.

To verify the ultraviolet renormalizability of the model and its 
infrared convergence, we have studied a set of exact results for 
different numbers $M$ and $N$. The same qualitative results as for 
$ M = - 21 $ and $ N = 16 $ can be obtained already for $ - M = N = 8$. 
We have checked the renormalizability by varying the ultraviolet limit 
$N$ from $4$ to $16$ and we verified the infrared convergence by varying 
the limit $M$ from $-4$ to $-21$. We have studied in detail the case 
$ - M = N = 12$. The eigenvalues in the latter case are almost the same 
as the corresponding eigenvalues in the case $ M = -21$ and $N = 16$ with 
the accuracy better than 1\% for the extreme eigenvalues. The intermediate 
eigenvalues match with much higher accuracy, correspondingly. For example, 
eigenvalues order 1 have the same 5 significant digits in both cases.

We calculate effective hamiltonians using the similarity renormalization 
group equations in the differential form. The effective hamiltonians 
are parametrized by their width in energy. The width is denoted by $\lambda$. 

We use Wegner's flow equation \cite{WEG} which provides a very elegant 
definition of the similarity transformation with a uniform band width 
and an explicit expression for the generating matrix. The effective 
hamiltonian matrices, ${\cal H} \equiv {\cal H}(s)$, are parametrized by 
the parameter $s$. $s$ ranges from $0$ to $\infty$. Wegner's original 
notation for this parameter was $\ell$ instead of $s$. It will be shown 
that the hamiltonian width is given by $\lambda = 1/\sqrt{s}$. 

The effective hamiltonian is divided into two parts,

$$ {\cal H} \quad = \quad {\cal D} + {\cal H} - {\cal D} \quad \equiv 
\quad {\cal D} + {\cal V} .
\eqno(2.5) $$

\noindent ${\cal D}$ is the diagonal part of the effective hamiltonian matrix, 

$$ {\cal D}_{mn} \quad = \quad {\cal D}_m \delta_{mn} , \eqno(2.6) $$

\noindent where ${\cal D}_m = {\cal H}_{mm}$. ${\cal V}$ is the 
effective interaction. It is equal to 
the off-diagonal part of the matrix ${\cal H}$. 
The Wegner flow equation is \cite{WEG}

$$ {d {\cal H} \over ds} \quad = \quad 
\left[\, [{\cal D}, {\cal H}],\, {\cal H}\, \right] ,
 \eqno(2.7) $$

\noindent with the initial condition 

$$ {\cal H}(0) \quad = \quad H \,\, . \eqno(2.8) $$

\noindent Equations (2.7) and (2.8) ensure that
${\cal H}(s)$ is a unitary transform of the initial Hamiltonian $H$.
In terms of the matrix elements, we have
 
$$
{d{\cal H}_{mn} \over ds} \quad = \quad -({\cal D}_m - {\cal D}_n)^2 
\, {\cal V}_{mn} \quad + \quad \sum_{\ell} ({\cal D}_m + {\cal D}_n 
- 2{\cal D}_\ell) \,
{\cal V}_{m \ell} \, {\cal V}_{\ell n} \,\, , \eqno(2.9)
$$

Equation (2.9) can be approximately solved for a  
small coupling constant $g$ by keeping only terms order $1$ and $g$. Namely,

$$
{\cal H}_{mn}(s) \quad = \quad E_m \, \delta_{mn} \quad - 
\quad g \sqrt{E_m E_n} \,\, 
\exp\left[ - s \, (E_m - E_n)^2 \right] \,\, . \eqno(2.10) $$

\noindent In this approximation, ${\cal D}_m = (1-g) E_m$. It is clear
that the parameter $s$ and the width $\lambda$ of the similarity 
renormalization scheme are simply related, $s = \lambda^{-2}$.
The similarity factor is a Gaussian function. 

It is well known that the coupling constant $g$ must depend on the upper 
energy bound, $2^N$, in order to eliminate the logarithmic divergences 
in higher orders when $N \rightarrow \infty$. Ref. \cite{WG} demonstrated 
that the Wegner flow equation has the standard renormalization group 
interpretation. Lowering the width $\lambda = s^{-1/2}$ 
in the discrete model is similar 
to lowering the upper bound on the energies of the interacting states. 
Therefore, in the higher order analysis, we replace
the coupling constant $g$ in Eq. (2.10) by the running coupling
$\tilde g (s)$ and use the expansion in 
$\tilde g (s)$ to remove the logarithmic divergences. We will gradually 
change the notation from the mathematical parameter $s$ to 
the hamiltonian energy width $\lambda$ in our parametrization of
the effective hamiltonians, including the running coupling constant.

The small coupling and large width approximation for 
$\tilde g (s)$ can be derived in the following way.
$E_m$ and $E_n$ are very small numbers for $m$ and $n$ close to $M$.
Therefore, Eq. (2.9) for $m$ and $n$ close to $M$, $g \ll 1$ and 
$s$ close to 0, reduces to

$$ {d \tilde g \over ds} \quad = 
\quad - \,\, \tilde g^2 \, {d \over ds} \sum_{\ell} 
\, \exp{[-2 \, {E}_{\ell}^2 \, s]} \,\, . \eqno (2.11) $$

\noindent Integration of Eq. (2.11) for our choice of the model parameters
gives the approximate running coupling constant

$$ \tilde g_a (\lambda) \quad = \quad 
{ g \over 1 - g \, (17.4 - \log{\lambda^2}/ \log{4})} \,\,
. \eqno(2.12)$$

\noindent $\lambda$ is taken in units of GeV. 
The number $0.4$ results from Eq. (2.11), because
the contributions of terms with $\ell$ close to 
$ \log{\lambda^2}/ \log{4}$ are smaller than $1$. 

Eq. (2.12) implies that the family of the effective theories exhibits typical 
asymptotic freedom behavior: the coupling gets smaller when the effective 
cutoff $\lambda$ is large.

The effective hamiltonians can be written now as

$$ {\cal H}_{mn}(\lambda) \, = \, E_m \delta_{mn} \, - 
\, \tilde g_a(\lambda) \,  
\sqrt{E_m E_n} \,\, \exp\left[ - [E_m - E_n]^2/\lambda^2 \right] 
+ corrections \,\, . \eqno(2.13) $$

\noindent Here, ${\cal D}_m(\lambda) = [1 - \tilde g_a(\lambda)] E_m 
+ corrections $. 

Eq. (2.13) demonstrates the utility of Wegner's equation. However, 
the uniform width of the effective hamiltonian distinguishes this 
solution from the widening band structure in the similarity scheme
from Refs. \cite{GW1} and \cite{GW2}. 
The widening of the band is useful in high 
order perturbation theory. A whole class of generalized Wegner
equations for the hamiltonian matrix elements can be written which
allow the high energy widening of the effective hamiltonian width. Namely,  

$$ {d {\cal H}_\lambda \over d\lambda^2} \quad  = \quad {- 1\over \lambda^4} 
\,\, \left[ F\{{\cal H}\}, \, {\cal H}_{\lambda} \right]
\,\, . \eqno(2.14)$$

\noindent In the original Wegner case, 
$F\{{\cal H}\} = [{\cal D}, {\cal H}]$. We have made 
calculations in the model using different
formulae for the operation $F$. For example, we used
$[F\{{\cal H}\}]_{mn} = \theta[|\Delta_{mn}| - x] 
(|\Delta_{mn}| - x)^k \Delta_{mn} {\cal H}_{mn}$,
 where $\Delta_{mn} = ({\cal D}_m - {\cal D}_n)/({\cal D}_m + {\cal D}_n)$
and $x$ is a function of $\lambda$ such that
$ 1 > x(\lambda) > x_0 > 0$, (see also Ref. \cite{G}). 
Nevertheless, we limit the present
paper discussion to the application of the original Wegner equation,
for simplicity and because no clear advantage of the more general equations
over the Wegner equation has been visible in the numerical studies we 
performed so far. 

The structure of Eq. (2.10)  demonstrates that the approximate solution 
including the running coupling cannot be obtained in the first order
perturbation theory. Eq. (2.11) shows that the running coupling  
constant approximation can be obtained in perturbation theory if one keeps 
terms order $\tilde g^2$. The approximate
running coupling $\tilde g_a (\lambda)$ is given in Eq. (2.12). 

The question is how large are the corrections indicated in Eq. (2.13).
The next Section describes our numerical study of the model.
\newpage
{\bf 3. NUMERICAL RESULTS}
\vskip.1in

The goal of our numerical study is to understand the structure and
determine the usefulness of the effective hamiltonians for the calculation
of the bound state energy.
We want to find out if perturbatively calculated effective hamiltonians 
of small width (of order 1 GeV) 
can reproduce the bound state eigenvalue. We are interested
in perturbation theory because it is the tool we can use 
in the hamiltonian approach to QCD. We aim to make the width $\lambda$
as small as possible
because the smaller is the width the smaller basis of 
effective states is
required to calculate the bound state eigenvalue.

The questions how small can be the width of the perturbatively calculated 
effective hamiltonians and how well the hamiltonians can reproduce the 
bound state properties, are of principal interest because the perturbative 
running coupling constant grows in asymptotically free theories when 
the width gets smaller and at some unknown point perturbation theory 
becomes useless. 

The approximate result for the running coupling constant
$\tilde g_a(\lambda)$ in Eq. (2.13) 
can be compared to the exact value of the running 
coupling. The exact solution is obtained by the numerical 
integration of the flow equation, i.e. Eq. (2.7). The numerical computations 
were performed using the Runge-Kutta integration algorithm of 
rank 4. \cite{NUM} The results were obtained using 
two independent alghorithms. The hamiltonians were  
cross-checked using various theoretical conditions such as 
the hermiticity and width-independence of the eigenvalues.

The exact running coupling, $\tilde g(\lambda)$, is defined as 

$$ \tilde g(\lambda) \quad = \quad - \,\, {\cal H}_{M, M+1}
(\lambda)/\sqrt{E_M E_{M+1}} \quad . \eqno(3.1) $$

Fig. 1. shows the running coupling constants $\tilde g$ and 
$\tilde g_a$ as functions of the width $\lambda$. It is visible that the
approximate solution blows up in the flow before the effective 
hamiltonian width is reduced to the scale where the bound state is formed. 

The bound state formation scale is defined using the relevant effective 
hamiltonian matrix element $\tilde \mu$ which is defined as 

$$  \tilde \mu (\lambda)\quad = \quad {\cal H}_{-1,-1}(\lambda)
\quad - \quad 0.5 \,{\rm GeV} \quad . \eqno(3.2) $$

\noindent $\tilde \mu$ becomes equal $-1.5$ GeV when the diagonal matrix 
element ${\cal H}_{-1,-1}$ of the effective hamiltonian becomes equal to the 
bound state eigenvalue, $-1$ GeV.  
The width $\lambda$ where the bound state eigenvalue appears on the diagonal 
is equal about $0.5$ GeV. $\tilde \mu$ stays constant for smaller 
$\lambda$. This width scale (order $1$ GeV) is called 
the bound state formation scale.  The exact result of integrating 
Eq. (2.7) gives the function $\tilde \mu(\lambda)$
which is plotted in Fig. 1. 

Fig. 1 also shows that the exact effective coupling constant is
close to the one in Eq. (2.13).
This is a hint for that the corrections indicated in Eq. (2.13) are small for 
$\lambda$ larger than order $16$ GeV. The effective
hamiltonians for smaller $\lambda$ may still have similar 
structure but the running coupling is not given by Eq. (2.12). 

The most important feature visible in Fig. 1 is that the exact 
effective coupling constant does not grow unlimitedly. This encourages
us to use expansion in the running coupling constant in calculations 
of the effective hamiltonians. 

On the other hand, 
the diagonal matrix elements ${\cal D}_m = (1- \tilde g) E_m$ 
with small $m$ become
negative when $\tilde g$ grows above $1$. \cite{WG} 
Thus, the absolute energy order of states is reversed. The diagonal 
matrix elements for
states corresponding to lowest $E_m$ become negative but they are
small in modulus. At the same time, the diagonal 
matrix elements for states corresponding to 
larger $E_m$ become more negative. Therefore, when $\tilde g$ grows above
$1$, the states corresponding to some originally intermediate 
energy range become much lower on the energy scale 
than all other states.

This inversion feature reverses the r\^ole of the bilinear terms. The coupling
stops growing and it begins to drop. At the same time the bound state 
eigenvalue is localized in the diagonal matrix element
${\cal H}_{-1,-1}$. This matrix element corresponds to the state 
which appears to be at the bottom
of the energy scale. The bound state dynamics is located in matrix elements
at the center of the effective hamiltonian matrix instead of the lowest 
indices corner, as one might expect if the lowest momentum scales 
were important in the bound state formation. Further renormalization 
group flow reduces $\tilde g$ below $1$ and successively 
establishes small positive eigenvalues on the diagonal. In other words,
the negative bound state eigenvalue is settled before many low positive 
eigenvalues are. The bound state eigenvalue appears almost independently 
of what happens at the bottom of the positive spectrum. This can be verified 
by changing $M$. The hamiltonian width $\lambda$ limits couplings
of a given state with other states to a limited number of states whose
diagonal matrix elements are within the range order $\lambda$ around the given
state diagonal matrix element. For $b^2 = 2$ only a small number of states
participate in the interaction of states with the diagonal matrix elements
larger than $\lambda$. Therefore, when $\lambda$ drops below the bound 
state formation scale, the small energy (small diagonal matrix element) 
states are decoupled from the bound state dynamics.

The rise of the coupling above 1 and the inversion feature 
suggest that even in the full hamiltonian flow the
perturbative expansion is useless at the bound
state formation scale. But the coupling approaches 1 from below
rather smoothly. We can ask for how small $\lambda$ the effective 
hamiltonian can be reliably calculated in perturbation theory. 

Scaling symmetry in the model and Fig. 1
suggest a large range of validity of perturbation theory
down to at least the width order $16$ GeV where the formula (2.12) 
begins to fail. Namely, the hamiltonians in Eq. (2.13) possess the discrete 
scaling symmetry: two effective hamiltonians for two values of 
the width $\lambda$ differing by 
the factor $b^{-2}$ have matrix elements related 
by the shift $i \rightarrow i - 1$ and $j \rightarrow j - 1$ 
and replacement of $\tilde g (\lambda)$ by $\tilde g (b^{-2} \lambda)$.
Thus, the flow of the effective hamiltonians is reduced to the change
of the width and the coupling constant for as long as one can neglect 
the boundary effects due to finite $N$ and $M$. Since the coupling 
in Eq. (2.12) is obtained keeping terms order $\tilde g^2$ and it runs
correctly down to about
$16$ GeV, one can expect that Eq. (2.13) describes the effective
hamiltonians correctly down to that scale. 

Fig. 2 shows the bound state eigenvalue of approximate
effective hamiltonians in ratio to the exact value $-1$ GeV, 
as a function of the width $\lambda$. Ratios formed this way 
will be used as measures of the accuracy of effective hamiltonians
throught this work.

Three approximations are shown in Fig. 2. The one denoted by $E$ results
from diagonalization of hamiltonians given by Eq. (2.13) with corrections
set equal 0 and inserting the exact value of $\tilde g (\lambda)$ 
for $\tilde g_a(\lambda)$. The curve labeled by $\cal D$ is obtained in the
same way except for one modification that in the exponent $E_m$ and
$E_n$ are multiplied by the factor $1 - \tilde g (\lambda)$. This
multiplication replaces $E_m$ and $E_n$ in the exponent 
by ${\cal D}_m$ and ${\cal D}_n$, respectively. 
The intermediate curve marked $g$ is obtained by 
using  ${\cal D} = (1 - g) E$ in the exponent, i.e. the initial
coupling instead of the running one.

The three curves show strong deviation 
from the exact eigenvalue at the beginning of the flow. This is caused by 
the boundary effect due to the cutoff $N$. The curves also deviate from 1
at small width. This deviation shows the limited validity of the 
approximation.
It is also clear that the gaussian similarity factor with free energies
gives a better approximation than the same factor with the diagonal matrix
elements. 

Fig. 2 suggests that a few terms of expansion
in the running coupling (e.g. only terms order 1 and $\tilde g$ are 
used in the case of curve $E$)
may produce effective hamiltonians whose bound state eigenvalue accuracy 
is order 10\% and whose width is smaller than $16$ GeV. The amazing result is
that by simply introducing the similarity function and varying the coupling 
one can reduce the width of the hamiltonian by a factor order $4000$ and make
a small error. Other eigenvalues are more accurate with the 
exception of eigenvalues order or larger than $\lambda$ for which the 
approximation is not expected to work.
However, the width is still quite far from the bound state formation scale.
The key question now is how far down in the width one can get using
an expansion in the running coupling constant.

The first comment we wish to make concerning perturbation theory 
is that the direct expansion in powers of the initial coupling $g$ 
is not useful for calculating effective hamiltonians with small width.
This is illustrated
in Fig. 3 by the plot of the ratio of the effective hamiltonian
bound state eigenvalue to the exact result for hamiltonians of width
$\lambda$ calculated using expansion into a series of powers of $g$ 
up to $1$, $2$, $3$ and $4$. Note that the analogy
between the model and QCD is such that $g$ in the model
corresponds to the strength of the two-particle interaction 
which is of the second 
order in QCD. Therefore, terms order $g$ here correspond to terms
order $g_{QCD}^2$, terms order $g^2$ correspond to terms order $g_{QCD}^4$ etc.
Fig. 3 clearly demonstrates that perturbative expansions in terms of the
canonical coupling constant in the initial hamiltonian are not suitable 
for applications in the bound state dynamics. 

The reason we stress this fact here despite that such result might 
be expected in the lagrangian approach, is that we study a hamiltonian 
approach. The hamiltonian approach 
is different in many respects from the familiar perturbative 
lagrangian approaches. \cite{W3} 
In the light-front hamiltonian approaches, it is 
often silently assumed that one can analyse nonperturbative aspects of 
the light-front QCD hamiltonian dynamics using canonical hamiltonian terms and 
neglecting the running coupling effects. Fig. 3 warns us not to do so.

In contrast to Fig. 3, Fig. 4 shows results  
obtained from effective hamiltonians calculated in the second order expansion 
in terms of the effective coupling constant $\tilde g(\lambda_0)$, 
for a set of choices of $\lambda_0$. The curves are marked by the value of 
$\lambda_0$ in GeV. For comparison, we include the second order result
of the expansion in terms of $g$ which is marked by the sign $\infty$
and equals to the curve marked 2 in Fig. 3.
The actual value of $\tilde g$ which is used to evaluate
the hamiltonians ${\cal H}(\lambda)$ is equal to the exact value of 
$\tilde g(\lambda_0)$ in the model. In a theory where an exact
solution is not known, the exact coupling constant $\tilde g(\lambda_0)$
is not known and
its value must be fitted in ${\cal H}(\lambda)$ at a useful value 
of $\lambda$. Varying of $\lambda$ should not cause significant changes
if the approximation and the fit are nearly exact.

The expansion of the renormalization group flow in powers of an 
effective coupling 
$\tilde g(\lambda_0)$ is done in the following way. One expands the 
flow in powers of the initial $g$ and computes $\tilde g(\lambda_0)$
as a series in $g$ using Eq. (3.1). The latter series is inverted and
$g$ is calculated as a series in $\tilde g(\lambda_0)$. Then, the whole
family of effective hamiltonians parametrized by $\lambda$ is calculated 
in expansion of powers of $\tilde g(\lambda_0)$ by substituting the inverted 
series into the known expansion in powers of $g$. 

Thus, the effective hamiltonians are calculated using expansion in powers of
$\tilde g(\lambda_0)$. They are diagonalized and their bound state
eigenvalues are plotted as functions of $\lambda$, for every single
value of $\lambda_0$ we choose. It is expected that an expansion in powers
of $\tilde g(\lambda_0)$ works only for effective hamiltonians with
widths $\lambda$ in the vicinity of $\lambda_0$. This feature is clearly
visible in Fig. 4 for $\lambda_0 = \infty$, $512$, $32$, $8$, $2$ and $1$ GeV. 
The arrows in Fig. 4 show points where $\lambda = \lambda_0$. 

The next question is what happens in the higher orders of 
perturbation theory. This question can be only partially answered on the
basis of the limited numerical studies we have performed. Therefore, we 
limit the discussion here to the analysis of the lowest orders in the
perturbative expansion for the key matrix element of Eq. (3.2) and
the bound state eigenvalue.

Firstly, note that Eq. (2.12) can be viewed as a result of 
summing a geometric series. It is clear that when $\lambda$ approaches 2 GeV, 
the coefficients in the expansion of $\tilde g_a$ in powers of $g$ 
grow as powers of $N$. Therefore, one should expect that when $g > 1/N$ 
the series will not converge. The value of $g$ we are using in the model
is only a little bit smaller than $1/16$, as required by the condition
for the bound state energy to be $-1$ GeV. For $\lambda \gg 2$ GeV the
coefficients are powers of much smaller numbers than $N$ and one can 
expect a small number of terms to reproduce the running coupling value, 
but when the width gets small, the number of terms in the expansion 
must become very large to reach the approximate result. When one switches to 
the expansion in the exact running coupling constant, the number of needed
terms is unknown. We will discuss the lowest 6 terms only.

Secondly, it follows from Fig. 1 that the exact solution for 
$\tilde g(\lambda)$ is limited and always stays below 1.1, 
in a dramatic distinction from the approximate running coupling 
result which diverges below $4$ GeV.
This suggests that the perturbative expansion for the full hamiltonian
may be extendible beyond
the barrier around 4 GeV. We show below it does happen so for the first
three terms in the expansion.


The difficulty we encounter with terms of higher order than $\tilde g ^2$
can be seen still using  
Eq. (2.12). Suppose $\lambda = 16$ GeV. We see in Fig. 1 that 
the second order perturbative result for $\tilde g_a(16 \,{\rm GeV})$ 
is quite close to the exact coupling constant $\tilde g(16 \,{\rm GeV})$. 
So, let us approximate the function $\tilde g (\lambda)$ by 
the function $\tilde g_a(\lambda)$ for $\lambda \geq 16$ GeV. Then,
using Eq. (2.12) at $\lambda = 16$ GeV, one obtains the inverse series

$$ g \quad = \quad \tilde g\, - \, (13.4 \, \tilde g)^2 \, +
\, (13.4 \, \tilde g)^3 \, - \, (13.4 \, \tilde g)^4 + ...
\,\, ,\eqno(3.3) $$

\noindent with the alternating sign. The first 6 actual coefficients 
which replace the successive powers 
of 13.4 in the full perturbative expansion are:  
$1.0$, $-13.42$, $177.3$, $- 2309$, $29752$ and $-379277$.
For example, the approximation in Eq. (3.3) deviates from the actual 
coefficients by about 5\% for the coefficient of $\tilde g^4$ 
and about 20\% for $\tilde g^6$.

Now, let us calculate the matrix element $\tilde \mu (16 \,{\rm GeV})$
of Eq. (3.2) as a power series in $g$. The actual coefficients are
$- 0.50$, $-6.71$, $-88.6$, $-1153$, $-14832$ and $-188759$. 
Inserting the expansion
from Eq. (3.3) one obtains the coefficients in $\tilde \mu (16 \,{\rm GeV})$
in front of the powers of $\tilde g(16 \,{\rm GeV})$, from the 
1st to 6th, equal $-0.5$, $0.0008$, $0.04$,
 $-0.14$, $1.2$ and $-17.3$, correspondingly.
One can easily test the series and see that it provides a good 
convergent approximation to the $\tilde \mu (\tilde g)$ for very small
values of $\tilde g$, such as 0.001 or even 0.01.
 
The exact value of the running coupling constant,
$\tilde g(16 \,{\rm GeV})$, is of the order 0.27. For this value
of $\tilde g$ one obtains the result 
$\tilde \mu (16 \,{\rm GeV}) = [- 0.13500 + 0.000057 + 0.00079 - 0.00074 + 
0.00174 - .00671]$ GeV. This sum does not include terms order 1. The 
first term comes from the order $\tilde g$, second from $\tilde g^2$ etc..

It is clear now that for the first three terms order $\tilde g^k$ with
$k = 0$, $1$ and $2$ we can expect convergence of the effective hamiltonian 
calculation while the higher orders may destroy it if one keeps only a 
few terms. 

The need for many orders in the expansion to reach convergence 
{\it beyond the first three terms} in the expansion can be suggested 
already on the basis of Fig. 1. Namely, we see that the
perturbative switching of $g_a(\lambda)$ from $+ \infty$ to $- \infty$
is smoothed out in the full renormalization group flow of $\tilde g(\lambda)$.
However, the turn of the curve $\tilde g$ below 1 GeV is sharp and
a large number of terms is required to reproduce the whole curve.

It is also visible, that the successive terms may cancel out. For example, 
the fourth order term may remove a large contribution of the third order term. 
If the numerical cancelation appears for many digits, the 
round off errors require a close examination. 

We have not performed extended studies of the model beyond the lowest order
terms. Therefore, we limit ourself to the presentation of Figs. 5 and 6 which 
illustrate what happens up to terms order $\tilde g^4$.
The effective hamiltonian ${\cal H}(\lambda)$
is expanded in a series of powers of $\tilde g(\lambda)$ (i.e. the 
running coupling at the same value of $\lambda$). 

Fig. 5 shows the accuracy of the perturbatively calculated 
effective hamiltonians as measured by the matrix element
$\tilde \mu(\lambda)$ from Eq. (3.2) in ratio to the exact value.
The expansion including terms order 1 and $\tilde g$
works with 10\% or even better accuracy down to the width 1 GeV.
It is also visible, that the expansion including
terms order $\tilde g^2$ produces accuracy on the order of
3\% or better down to the width order 2 GeV but the accuracy 
drops down significantly near 1 GeV. The third power term
produces a considerable drop in the accuracy which is then 
partially counterbalanced by the inclusion of the fourth power term.
The oscillation pattern needs further studies to explain.
We have not performed such studies.

Fig. 6. shows the accuracy of the bound state eigenvalue obtained from 
diagonalization of the effective hamiltonians ${\cal H}(\lambda)$
expanded in powers of the running coupling $\tilde g (\lambda)$ 
including powers up to 1, 2, 3 and 4, respectively. The curve 1 for
the first order calculation
matches curve $E$ in Fig. 2 away from the ultraviolet (left) boundary region.
This feature confirms the expectation based on Figs. 1 and 2 that 
a low order perturbative expansion in the running coupling constant 
can accurately reproduce the effective
hamiltonians at small width. The curve 2 represents the result 
of the second order expansion and shows a considerable improvement.
The accuracy is about 10\% or better down to the width $1$ GeV which is at the
edge of the bound state formation scale.
The second power of $\tilde g$ in the model is analogous to the 
fourth power of the coupling constant in QCD. Thus, the analogous calculation
in QCD requires the fourth order expansion of the effective hamiltonians 
in powers of the strong interaction coupling constant.

It is visible in Fig. 6 that for the small values of $\tilde g$ the 
perturbative expansion is convergent. However, as expected, at small widths 
the powers 3 and 4 
of the coupling constant appear with large coefficients. The curve
including terms up to the third power of $\tilde g(\lambda)$ 
falls down quite low
when we substitute the exact value of the coupling.

The fourth order curve labeled 4 is ended at the bottom edge of the figure to
avoid overlap with curves 1 and 2. In fact, the curve 4
continues down to about 0.53 at $\lambda = 16$ GeV, shoots back up
to 0.94 at about 12 GeV, deeps down again to about 0.5 at 8 GeV and 
skyrockets right after 8 GeV crossing 1 and reaching about 100
at 1 GeV. This erratic behavior is clearly correlated with the pattern
visible in Fig. 5. 

Despite the slow convergence problem which requires further study,
Figs. 1 to 6 illustrate the striking feature of the model that
the bound state eigenvalue can be obtained from diagonalization of
an effective hamiltonian with the width order 1 GeV with 10\% accuracy.

The remaining question is how small the space of states can be
on which one can project the narrow effective hamiltonian and reproduce
the bound state eigenvalue by diagonalization of the projected matrix. 
The wave function is expected to be reproduced with a similar accuracy.

The answer is provided in Table 1. The table contains the ratio 
of the bound state eigenvalue obtained by the diagonalization of
a small window matrix $\tilde {\cal H}(\lambda)$ whose matrix 
elements are the same 
as the matrix elements of ${\cal H}(\lambda)$ in the small window
but they equal 0 everywhere outside the window. The indices of the window 
range from $\tilde m$ to $\tilde n$ including the limiting values.
The bound state eigenvalue of the window is divided by the bound state
eigenvalue of the whole effective hamiltonian. Table 1 gives the results
obtained from $\tilde {\cal H}(1 \,{\rm GeV})$ which is calculated 
in second order expansion in the running
coupling constant $\tilde g(1 \,{\rm GeV})$. The bound state eigenvalue 
of the whole 
effective hamiltonian ${\cal H}(1 \,{\rm GeV})$ is equal $ - 0.8902$ GeV
(the exact value is $ - 1 $ GeV). The entries in Table 1 show that the small
window matrix easily reproduces this result with a relatively high accuracy.
This is a spectacular feature
of the method and the model. For the coupling constant 
$\tilde g(1 \,{\rm GeV})$ is equal $1.05$ and
still the bound state eigenvalue accuracy one obtains from the small
window hamiltonians calculated in second order
perturbation theory is on the order of 10\%.

Table 1 demonstrates that our hamiltonian approach can be used
to calculate a $10 \times 10$ or even $5 \times 5$ matrix 
whose lowest eigenvalue
reproduces the full theory bound state eigenvalue with accuracy
order 10 to 20\%. This model result can be viewed as encouraging
to pursue a similar strategy in case of QCD. However, the 
slow down or absence of convergence beyond the second order 
expansion near the bound state formation scale,
require improvements.

\vskip.5in
{\bf 4. CONCLUSION}
\vskip.1in

We have studied two basic approximations which may be of help in the
hamiltonian calculations of bound state properties in asymptotically
free theories. The first one can be briefly described as constructing
effective hamiltonians by introducing
the similarity factors and adjusting coupling constants. 
We provide the definition of the asymptotically free running coupling
constant in the hamiltonian approach. The second approximation is 
an expansion into a series of powers 
of the running coupling constant. In both cases,
the bound state eigenvalues and eigenstates of the full theory are 
found by diagonalizing the effective hamiltonians.

Our model study suggests that one may hope to calculate
effective hamiltonians down to the similarity width which is close to the
bound state formation scale. The coupling constant growth is limited.
Moreover, the effective hamiltonian can be diagonalized in a limited subspace 
of states which dominate the bound state dynamics, instead of using the full 
basis. The small window hamiltonian reproduces the whole effective hamiltonian
bound state eigenvalue with accuracy order 10\% or better.

The model exhibits an inversion of the energy hierarchy of states when 
the coupling constant becomes slightly larger than 1. On the one hand, 
this feature is welcomed because it decouples the small momentum scales
from the bound state dynamics. On the other hand, this feature appears at the
coupling larger than 1 and it is beyond the reach of a simple 
perturbation theory. 
One can ask if a similar difficulty must exist in QCD. The expected answer 
is no. For one may hope that gluons effectively obtain masses already
at small values of the coupling constant through the 
gluonic couplings. These masses (or potentials) may set the states with gluons
appart in energy when the width gets small before the coupling becomes
too large for being treated in perturbation theory.

A comment is in order concerning the similarity approach
in view of the latest findings in the lattice calculations
that glueballs are unexpectedly heavy. \cite{Lepage} 
The hypothesis is that glueballs are heavy
because the effective potentials binding two gluons are much stronger
(perhaps by a factor of four) than the potentials that bind quarks.
The reason is that gluons are in the octet representation of SU(3),
which is analogous to being a doubly charged object in SU(1). Thus, if
one compares a quark bound state to hydrogen, a glueball bound state
is comparable to helium, but with one doubly-charged electron rather
than two singly charged electrons. Obviously the helium and doubly
charged electron are far more strongly bound than hydrogen is.

	What this means is that as the coupling constant increases with
decreasing hamiltonian width, the coupling of gluons will always be four times
larger than the coupling for quarks, which suggests the gluon coupling
would be expected to become strong enough to create a bound state of
gluons while still too small to bind quarks. The strong gluon binding
would naturally lead to a high effective mass for gluons, making them
unlikely to be present in quark bound states. This would explain why
we see no evidence of explicit constituent bound states involving
gluons as well as quarks, yet allow the quark-quark potential to have
strong higher order corrections. A major question for this picture
would be to understand sum rules for deep inelastic scattering which
have been interpreted as implying a large contribution from
constituent gluons inside the proton. It might be that development of
the similarity transformation formalism would show that sum rule data
refers more to "current" quarks and gluons, before the similarity
transformation is applied, rather than to constituent quarks and
gluons after the transformation, where the gluon contribution could be
very small by these arguments.

Our study shows that the convergence of perturbation theory
is in jeopardy beyond the second order expansion for large 
values of the running coupling constant.
Ref. \cite{W3} suggested that one can use the phenomenological 
success of the constituent quark model to improve convergence 
when solving QCD in the light-front hamiltonian approach. 
The improvement is expected to result from using a constituent quark
hamiltonian as a first approximation to the small width effective
hamiltonian of QCD. The chance exists, that such a constituent 
quark hamiltonian is not much different from the
theoretical one in QCD. Therefore, the distance to the true effective
hamiltonian may be calculable in perturbation theory. 
We wish to add a comment on how this suggestion could be checked numerically
in the model.

The analogous step could be done in the present model in the following 
way. For certain $\lambda_0$ close to the bound state formation scale,
order a few GeV, the running coupling, $\tilde g_0 \equiv
\tilde g(\lambda_0)$, has a value comparable to $0.5$. Let us denote 
the exact value of $\tilde g_0$ by $g_s$ ($s$ is chosen for ``strong'').
The perturbative expansions
towards smaller widths order $1$ GeV in terms of $\tilde g(\lambda)$
are hard to continue because of the large distance from the
small coupling $g$ in the initial hamiltonian.
In these circumstances, we can add and subtract a suitable term
in the hamiltonian ${\cal H}(\lambda_0)$, say $\mu_{CQM}$. This step changes
nothing. But we can multiply the subtracted term by the ratio 
$\tilde g_0 / g_s$. For $\tilde g_0 = g_s$ nothing is changed in the
theory. On the other hand, if we replace $\tilde g_0$ by a small number,
the subtracted term together with the original interaction can be treated
as a small perturbation. In fact, if $\mu_{CQM}$ represents the bulk of
the effective hamiltonian then the difference between the effective 
hamiltonian and $\mu_{CQM}$ will not lead to large corrections even if 
$\tilde g_0$ is rised to $g_s$. Thus, perturbation theory in terms of 
$\tilde g_0$ could be continued towards the smaller widths and the coefficients
could be kept small. One can think of the hamiltonian
$\mu_{CQM}$ as a matrix which has only one matrix element different
from 0, right in the place on the diagonal where the bound state eigenvalue
appears. Other forms are also possible. The key example is provided by 
the approximations shown in Fig. 2. The general feature of the example
is that it amounts to the insertion of the similarity factors and 
adjusting the couplings.

Another opportunity for improvement in the numerical accuracy is
related to the irrelevant operators. 
Analysis of the renormalization group equations for low energy 
matrix elements suggests that for small widths $\lambda$ the {\it corrections} 
on the right-hand side in Eq. (2.13) should include the term $h(\lambda)$
with matrix elements of the form

$$ h_{mn}(\lambda) \, = {\tilde h(\lambda)\over \lambda} \, (E_m + E_n) 
\sqrt{E_m E_n} \,\, \exp\left[ -[E_m - E_n]^2 /\lambda^2 \right] 
\,\, . \eqno(4.1) $$

\noindent A new coupling constant $\tilde h \sim \tilde g^2$ is introduced. 
The new term behaves as $\lambda^{-1}$ for large $\lambda$
and it disappears for $\lambda \rightarrow \infty$. 
It remains to be verified how much this term can improve the accuracy of 
the perturbative evaluation of narrow effective hamiltonians and the resulting
bound state eigenvalues. The two couplings $\tilde g$ and  $\tilde h$
are related and should be considered coherently. \cite{PW}

Too little 
is known yet about equations of the type 
(2.14) to say if they can help in accelerating convergence
of the perturbative expansion.

We should mention that when $b$ is reduced towards 1 the number 
of states per unit of the energy grows and the resulting matrices would have to
be much larger than in the example we described. In fact, an interesting
problem arises when one considers more than one state at each scale. Namely,
when we have only one state per scale then no degenerate or near neighbour 
interactions arise. If more states appear at each scale and they couple
to the near neighbours, the couplings are not reduced
by the similarity transformation. The interesting question is how large 
can be the effective energy range of such near neighbour interactions.

\vskip.3in
{\bf Acknowledgement}
\vskip.1in

One of us (SG) would like to thank The Ohio State Physics Department 
Nuclear, High Energy and Physics Education Groups, especially Robert 
Perry, Billy Jones, Martina Brisudov\'a and Brent Allen, for the 
hospitality and discussions during his stay at OSU as a Fulbright Scholar
in the academic year 1995/96 where some of the initial work on this model 
was done. Research described in this paper has been supported in part by Maria
Sk{\l}odowska-Curie Foundation under Grant No.  MEN/NSF-94-190.

\newpage

{\bf Figure Captions}

\vskip.2in

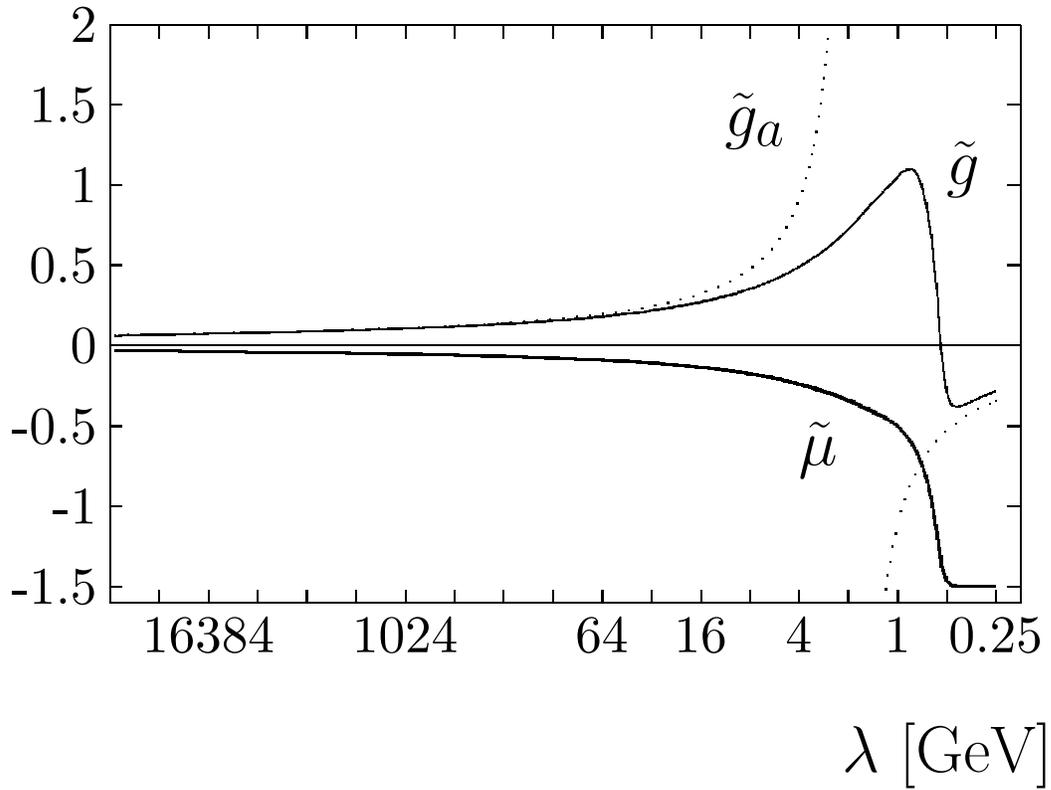
{\bf Fig. 1.} The approximate running coupling $\tilde g_a(\lambda)$ 
from Eq. (2.13) and the exact running coupling $\tilde g(\lambda)$ 
from Eq. (3.1) plotted as
functions of the effective hamiltonian width $\lambda$. The 
exact reduced matrix element $\tilde \mu$ from Eq. (3.2) is also plotted
to show the width range where the bound state eigenvalue appears on the
diagonal. 

\vskip.2in

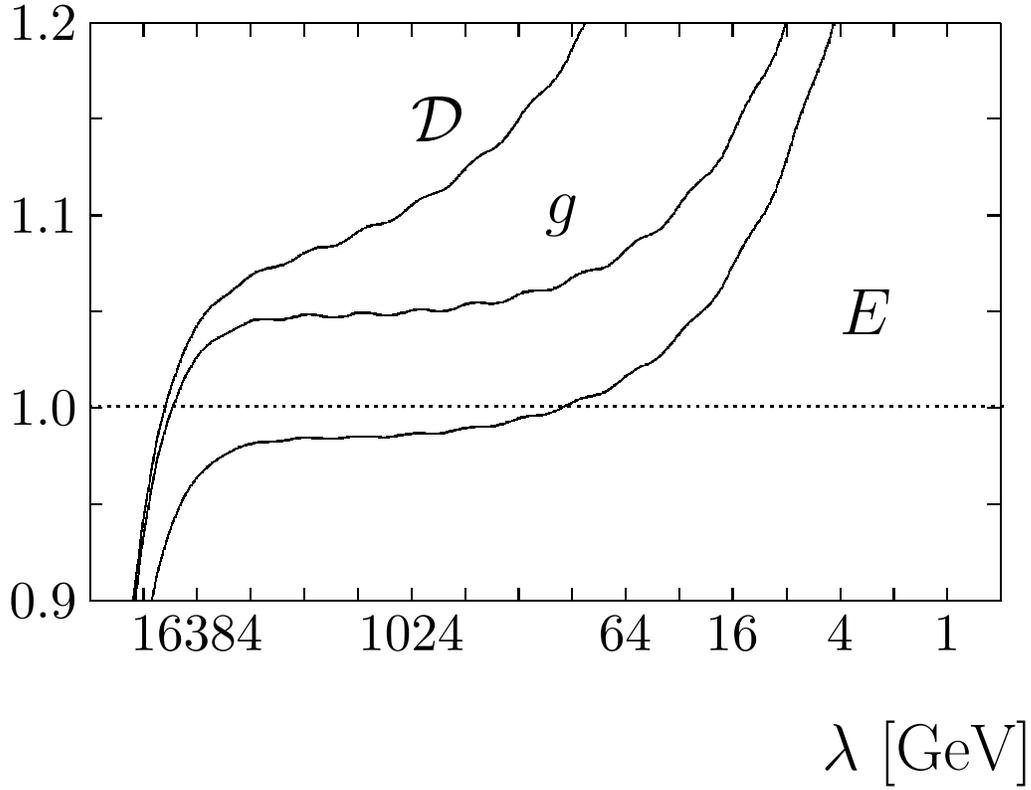
{\bf Fig. 2.} The bound state eigenvalue of approximate
effective hamiltonians in ratio to the exact value $-1$ GeV. 
Curve $E$ results from Eq. (2.13) with 
{\it corrections} set equal 0. Curve $\cal D$ is obtained in the 
same way as $E$ except for $H_0$ eigenvalues $E$ in the Gaussian 
similarity factor replaced by the diagonal matrix elements of the
approximate effective hamiltonian, i.e. ${\cal D} = (1-\tilde g) E$, 
(see the text). The intermediate curve marked $g$ is obtained by 
using  ${\cal D} = (1 - g) E$ in the exponent, i.e. the initial
coupling instead of the running one.

\vskip.2in

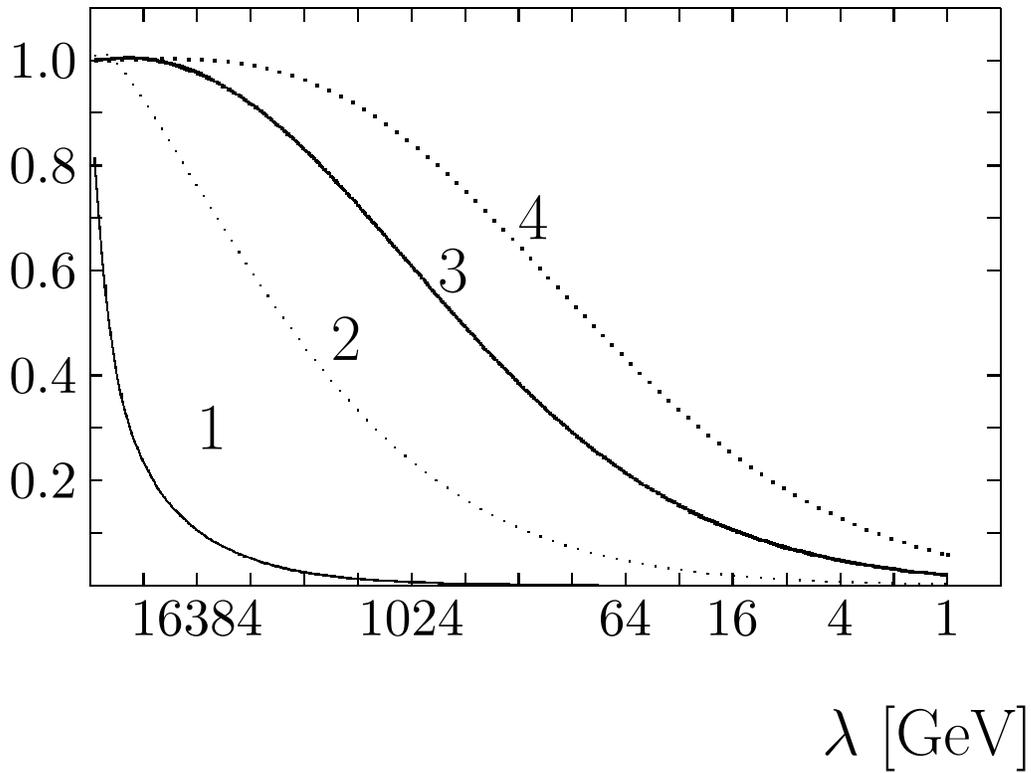
{\bf Fig. 3.} The ratio of the effective hamiltonian
bound state eigenvalue to the exact result, for the hamiltonians of width
$\lambda$ calculated using expansion in powers of the initial coupling
constant $g$ up to $1$, $2$, $3$ and $4$, respectively. It is visible, that
the expansion is not useful in the calculation of the bound state eigenvalue.

\vskip.2in

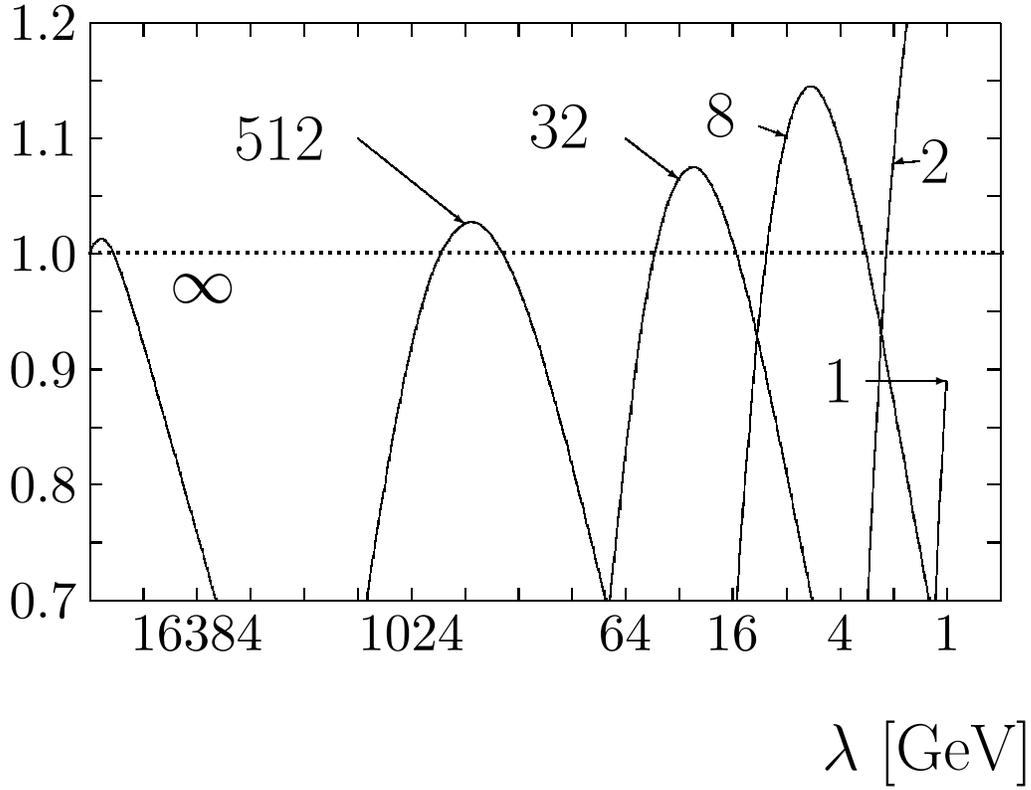
{\bf Fig. 4.} The accuracy of the bound state eigenvalues obtained
from effective hamiltonians whose renormalization group flow with
the width $\lambda$ is
calculated expanding in powers of the effective coupling constant
$\tilde g (\lambda_0)$ and including terms order $1$,
$\tilde g (\lambda_0)$ and $\tilde g^2 (\lambda_0)$. 
The accuracy is given as ratio of the
bound state eigenvalue obtained by diagonalization of the effective
hamiltonian of width $\lambda$ to the exact value, $- 1$ GeV.
The curves correspond to the indicated values of 
$\lambda_0$ (in units of GeV). The result of expansion in
the initial coupling $g$ is denoted by $\infty$. 
The arrows show points where $\lambda = \lambda_0$.

\vskip.2in

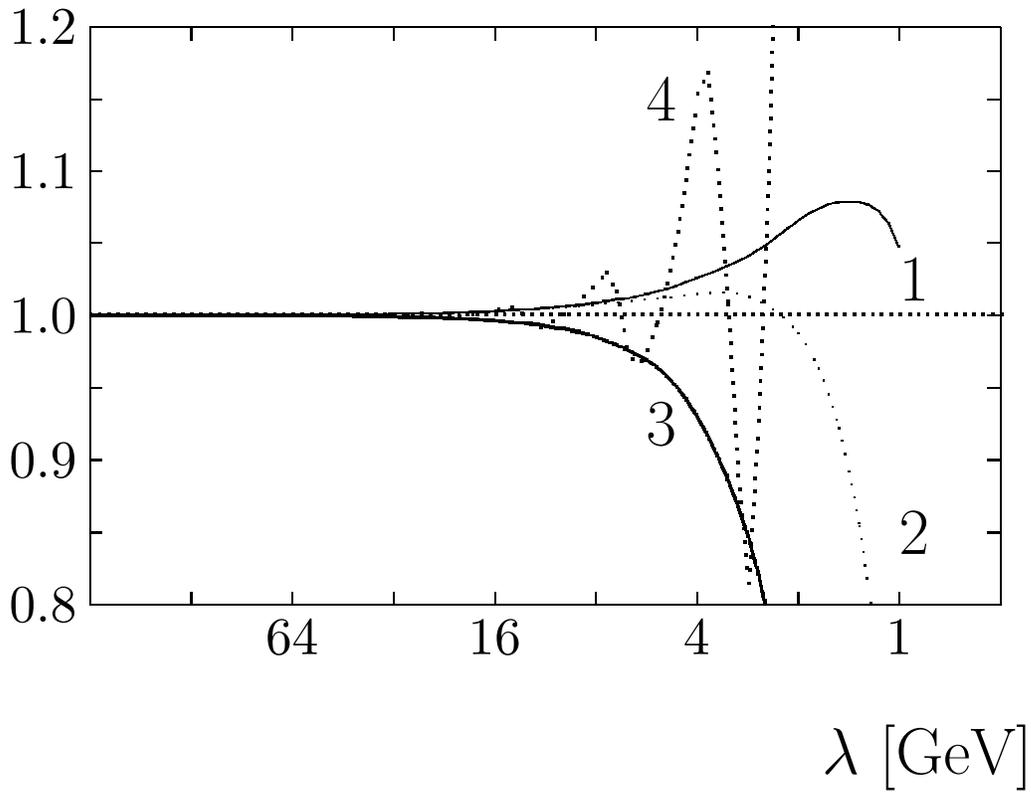
{\bf Fig. 5.} The accuracy of the perturbatively calculated 
effective hamiltonians as measured by the matrix element
$\tilde \mu(\lambda)$ from Eq. (3.2) in ratio to its exact value.
The numbers label results obtained by expanding in powers of the
running coupling constant $\tilde g (\lambda)$ including powers
up to 1, 2, 3 and 4, respectively. The ratio would be 
indistinguishable from 1 for widths larger than 64 GeV if it were
plotted in the scale of this figure.

\vskip.2in

{\bf Fig. 6.} Accuracy of the bound state eigenvalue obtained from 
diagonalization of the effective hamiltonians ${\cal H}(\lambda)$
expanded in powers of the running coupling $\tilde g (\lambda)$ 
including powers up to 1, 2, 3 and 4, respectively. The curve 1
matches curve $E$ in Fig. 2 away from the left boundary region.
The curve order 2 shows 10\% accuracy down to the width $1$ GeV.
The lack of convergence for higher order curves at small widths
is discussed in the text.

\vskip.2in

{\bf Table 1.} 
Ratio of the bound state eigenvalue of the small window hamiltonian 
with indices limited by $\tilde m$ and $\tilde n$, to the eigenvalue 
of the whole effective hamiltonian at $\lambda = 1$ GeV calculated using 
expansion up to second power in the running 
coupling $\tilde g(1 \,{\rm GeV})$. 
0.993 corresponds to the absolute accuracy of the bound state eigenvalue 
equal 12\% and 0.908 to 19\% (see the text).

\end{document}